\documentclass{elsart}
\usepackage{graphicx}
\usepackage{graphics}
\usepackage{epstopdf}
\usepackage{epsfig}
\usepackage{amssymb}
\usepackage{lineno}
\usepackage{bm}
\usepackage{pifont}
\usepackage{color}
\usepackage{marvosym}
\usepackage{enumerate}

\begin{document}

\begin{frontmatter}

\title{Hypernuclear single particle spectra based on in-medium chiral $SU(3)$ dynamics
}
\author[label2]{P. Finelli}
\ead{paolo.finelli@bo.infn.it},
\author[label3]{N. Kaiser},
\author[label4]{D. Vretenar},
\author[label3]{W. Weise}
\address[label2]{Physics Department, University of Bologna,
I-40126 (Italy)\\
INFN, section of Bologna}
\address[label3]{Physik-Department, Technische
  Universit\"at M\"unchen,\\ D-85747 Garching (Germany)}
\address[label4]{Physics Department, University of Zagreb, HR-10002 (Croatia)}

\begin{abstract}

A previously derived relativistic energy density functional for nuclei, 
based on low-energy in-medium chiral dynamics, is generalized to implement 
constraints from chiral $SU(3)$ effective field theory and applied to $\Lambda$ hypernuclei.
Density-dependent central and spin-orbit mean fields 
are calculated for a $\Lambda$ hyperon 
using the $SU(3)$ extension of in-medium chiral perturbation theory
to two-loop order. 
Long range $\Lambda N$ interactions arise from
kaon-exchange and from two-pion-exchange with a $\Sigma$ hyperon
in the intermediate state. Short-distance dynamics is encoded in contact
interactions. They include scalar and vector mean
fields reflecting in-medium changes of quark condensates, constrained by QCD sum rules.
The $\Lambda$ single particle orbitals are computed for a series of hypernuclei 
from $^{13}_{~\Lambda}$C to  $^{208}_{~~\Lambda}$Pb. 
The role of a surface (derivative) term is studied. Its strength is found to be compatible
with a corresponding estimate from in-medium chiral perturbation theory. 
Very good agreement with hypernuclear spectroscopic data is achieved. 
The smallness of the $\Lambda$-nuclear spin-orbit interaction
finds a natural explanation in terms of an almost complete
cancellation between short-range scalar/vector contributions and longer range
terms generated by two-pion exchange.

\end{abstract}


\journal{NPA}

\end{frontmatter}

\newpage

\section{\label{secI} Introduction} 
The spectroscopy of $\Lambda$ hypernuclei \cite{hypspinorbit, hypspinorbit2, hypspinorbit3} sets strong constraints \cite{Millener_rev} on the spin dependence of the $\Lambda N$ effective interaction. There is convincing evidence that the $\Lambda$-nuclear spin-orbit interaction is abnormally weak compared to the very strong spin-orbit force experienced by nucleons in ordinary nuclei. Experiments setting limits on the spin-orbit coupling in light $\Lambda$ hypernuclei are reported in Ref. \cite{Kohri:2001ncAjimura:2001naAkikawa:2002tm}. A recent analysis \cite{Motoba_08} of heavy hypernuclei confirms the systematic smallness of the $\Lambda$-nuclear spin-orbit splitting.

Many theoretical attempts were made over the years to understand the weakness of the 
$\Lambda$-nuclear spin-orbit coupling \cite{BrockmannWeise, noblepirner, Gal_insertion, jennings, tensor}, with model assumptions ranging from quark degrees of freedom via phenomenological boson exchange mechanisms to unusually strong (negative) $\omega\Lambda$ tensor couplings. The systematics of $\Lambda$-nuclear shell model orbitals have been studied in a variety of approaches. The quark-meson coupling model \cite{Tsushima_old,Tsushima_new} seeks the origin of the weak spin-orbit coupling  of the $\Lambda$ at the quark level. Relativistic mean-field (RMF) models  \cite{BrockmannWeise, Keil:1999hk} generate the spin-orbit coupling by the coherent interplay of scalar and vector mean fields. Even though these fields are only about half as strong for $\Lambda$ hyperons as compared to those for nucleons in nuclei, the resulting $\Lambda$-nuclear spin-orbit splittings are usually still too large unless additional ad-hoc mechanisms are invoked for their suppression. 
In mean-field descriptions with phenomenological density dependent interactions of Skyrme type \cite{MDG,Skyrme, SKdef}, and in several hypernuclear many-body calculations (Fermi Hypernetted Chain, Brueckner-Hartree-Fock) \cite{FHNC,BHF}, the spin-orbit splitting is suppressed simply by hand. 

In this article we follow a different path motivated by recent developments at the interface between low-energy QCD and the nuclear many-body problem. Chiral effective field theory is established as the  realization of QCD in the low-energy limit, with pions as Goldstone bosons of spontaneously broken chiral symmetry being the active light degrees of freedom. In its version with nucleons and $\Delta(1230)$ isobars as heavy baryons, chiral effective field theory is considered to be an appropriate starting point not only for describing low-energy pion-nucleon interactions but also for the nuclear many-body problem \cite{EHM,KFW}. 

In-medium chiral perturbation theory \cite{KFW,FKW} provides a successful  framework for constructing the energy density of nuclear matter as a function of Fermi momentum $k_F$. Long and intermediate range one- and two-pion exchange interactions including tensor and three-body forces are treated explicitly. Short-distance dynamics, not resolved in detail at the characteristic Fermi momentum scales, is encoded in contact terms. The contact terms generate contributions to the energy per particle that are linear in the density $\rho = 2\,k_F^3/(3\pi^2)$. These terms need to be adjusted, e.g. by reproducing the empirical binding energy, while terms of higher (fractional) powers in the density, involving effects of the filled Fermi sea on two-pion exchange processes, are computed with input fixed entirely by empirical low-energy pion-nucleon data. Three-nucleon interactions are systematically incorporated. Spin-orbit forces follow consistently from the evaluation of spin-dependent terms in inhomogeneous nuclear matter \cite{KFW:2003,Kaiser:2003,Kaiser:2003uh,Kaiser:2003xz,Kaiser:2002yg}.

The translation of this framework into a relativistic energy density functional for finite nuclei \cite{Finelli0304, Finelli0607} is designed so that it keeps the separation-of-scales concepts of the previous nuclear matter calculations. It includes explicit pion and two-pion exchange dynamics up to three-loop order in the energy density. It also incorporates strong scalar and vector mean fields (equivalent to contact interactions and constrained by in-medium QCD sum rules), and an additional surface (derivative) term. Applications of this approach to nuclei have been quite successful throughout the nuclear chart and naturally suggest a systematic extension to hypernuclei, with special focus on the very different spin-orbit interactions of nucleons and $\Lambda$ hyperons. 

The generalization to chiral SU(3) introduces a well-defined set of couplings of the pseudoscalar meson octet to the baryon octet. The longest range $\Lambda N$ interactions are now generated by two-pion exchange with an intermediate $\Sigma N$ and (less importantly) by kaon exchange. The even longer range one-pion exchange is excluded in lowest order by the isospin $I=0$ of the $\Lambda$. Contact interactions are again representing unresolved short-distance dynamics. A first study using this in-medium chiral SU(3) dynamics approach was performed for $^{16}_{~\Lambda}$O as a test case \cite{Finellihyper}. In the present work these calculations are systematically expanded over a large range of hypernuclei from $^{13}_{~\Lambda}$C to  $^{208}_{~~\Lambda}$Pb.

The basic mechanisms that govern spin-orbit interactions in comparison between nuclei and hypernuclei are a persistent theme of the present  study. Three major sources of spin-orbit interactions in nuclear systems can be identified: 

\begin{enumerate}[ (i)]
\item short-distance dynamics of coherently acting scalar and vector mean fields; 
\item intermediate range spin-orbit forces induced by the pion exchange tensor interaction in second order, with Pauli blocking of intermediate nucleon states;
\item a three-body  spin-orbit interaction of Fujita-Miyazawa type \cite{fujita}, produced by two-pion exchange with intermediate excitation of a virtual $\Delta$ isobar.
\end{enumerate}

All of these mechanisms are generated within the in-medium chiral dynamics framework. 
The important feature pointed out in Refs. \cite{KFW:2003,KWhyp} and further elaborated in Ref. \cite{KWproposal}, is that mechanism (ii) comes with opposite sign but similar magnitude as compared to mechanisms (i) and (iii). The balance between all three mechanisms is shown to account for the large residual spin-orbit splitting observed empirically in nuclei. For a $\Lambda$ in a hypernucleus, however, the three-body Fujita-Miyazawa mechanism (iii) has no analogue simply because the $\Lambda$ exists only as a single valence particle and there is no hyperon Fermi sea. This implies that the short-distance (or scalar-vector) mechanism (i) and the intermediate range, second-order tensor force mechanism (ii) largely cancel \cite{KWhyp} to make a small net spin-orbit splitting for the $\Lambda$-hypernuclear orbitals. Testing this scenario 
over a large set of hypernuclei is one of the primary goals of the present work.

\section{Theoretical framework}
\subsection{Hypernuclear energy density functional}
A reliable and accurate calculational framework to deal with fermionic many-body systems such as nuclei and hypernuclei is the density functional approach. Here we start from a relativistic energy density functional for hypernuclei with a single $\Lambda$ hyperon orbiting in the nuclear environment:
\begin{equation}
\label{E_lambda}
E[{\rho}]  =  E^N[{\rho}] + E_{\rm free}^\Lambda[{\rho}] 
+ E_{\rm int}^\Lambda[{\rho}]\,, 
\end{equation}
where $E^N$ is the energy of the nuclear core and
\begin{eqnarray}
\label{E_lambda_free}
E_{\rm free}^\Lambda & = & 
\int d^3r \langle \Phi_0 | \bar{\psi}_\Lambda [-i 
{\bm{\gamma}} \cdot {\bm{\nabla}} + M_\Lambda ]
\psi_\Lambda |\Phi_0 \rangle~, \\
\nonumber\\
\label{E_lambda_int}
E_{\rm int}^{\Lambda} & = & 
\int d^3r \left\{ \langle \Phi_0 | 
G^{\Lambda}_S ({\rho}) \left( \bar{\psi} \psi \right) \left(
\bar{\psi}_\Lambda \psi_\Lambda \right) | \Phi_0 \rangle~  \right. \nonumber\\
& ~ & \left. \quad \quad \quad + ~\langle \Phi_0 | G^{\Lambda}_{V} 
({\rho}) \left( \bar{\psi} \gamma_\mu \psi \right) \left(
\bar{\psi}_\Lambda \gamma^\mu \psi_\Lambda \right) | \Phi_0 \rangle~  \right. \nonumber \\
\label{der}& ~ & \left. \quad \quad \quad\quad + ~ \langle \Phi_0 | D^{\Lambda}_{S}
\,\, \partial_\mu ( \bar{\psi} \psi) 
\,\partial^\mu(\bar{\psi}_\Lambda \psi_\Lambda) | \Phi_0 \rangle
\right\} \; ,
\end{eqnarray}
are the additional contributions to the energy involving the hyperon. Here
$|\Phi_0 \rangle$ denotes the hypernuclear ground state; $\psi_\Lambda(x)$ and $\psi(x)$ are the hyperon and nucleon fields, respectively. 

The $E[\rho]$ of Eqs. (\ref{E_lambda}-\ref{E_lambda_int}) represents a generalization to hypernuclei 
of the nuclear energy density functional \cite{Finelli0304, Finelli0607}, constrained 
by basic features of low-energy QCD. The nuclear
part $E^N[{\rho}]$, introduced in Ref. \cite{Finelli0607}, 
describes the core of interacting protons and neutrons in terms of 
the corresponding isoscalar and isovector densities and currents\footnote{
We retain the notation adopted in Refs. \cite{Finelli0304,Finelli0607, Finellihyper}: space vectors are denoted with boldface characters (${\bm x}$), 
vectors in isospin space with an arrow ($\vec{\tau}$); Greek symbols are used
for space-time indices.}.
The hypernuclear functional, Eq. (\ref{E_lambda}), includes 
the $\Lambda$ kinetic energy and mass term $E_{\rm free}^\Lambda$ of 
Eq. (\ref{E_lambda_free}), and the term $E_{\rm int}^{\Lambda}$ of Eq. (\ref{E_lambda_int}) which summarizes  $\Lambda$-nucleon interactions in the nuclear environment. 

Also included in this density functional is a surface term proportional to $D_S^\Lambda$
that involves gradients of the isoscalar-scalar nucleon density and the scalar 
hyperon density distributions.  Such a term arises naturally in the gradient expansion
of the energy density functional for a finite system.  The leading pieces (the first two terms 
on the r.h.s. of
Eq. (\ref{E_lambda_int})) correspond to the local density approximation (LDA).
They account for the interaction of a $\Lambda$ with homogeneous isospin-symmetric
nuclear matter taken at the actual local density of the nuclear core. 
The $\Lambda$-hypernucleus is spatially inhomogeneous, however, and surface
effects are expected to be non-negligible. The gradient term (third term in  
Eq. (\ref{E_lambda_int})) is introduced
to account for such corrections beyond the LDA\footnote{The gradient term introduced here 
is a step beyond our previous hypernuclear study \cite{Finellihyper} which employed a simpler LDA model.}. The explicit form of this surface term
is guided by the corresponding part of $E^N[{\rho}]$ in the nucleon sector \cite{Finelli0607}.

\subsection{Effective interaction}
The strength of the effective $\Lambda$-nucleon interaction in the nuclear medium, see 
Eq.(\ref{E_lambda_int}), is determined by density-dependent vector
and scalar couplings, $G^{\Lambda}_V(\rho)$
and $G^{\Lambda}_S(\rho)$, of dimension (length)$^2$. We follow here a strategy analogous to our previous calculations for
nuclei. Long and intermediate range hypernuclear dynamics are governed by chiral two-pion and kaon exchanges in the presence of the nuclear Fermi sea. 
Short distance dynamics is encoded in contact terms which generate Hartree type mean-field contributions linear in the density. The hyperon self-energies resulting from both long and short range interactions are then transcribed into equivalent density dependent couplings, expanded in (fractional) powers of the local density $\rho(\bm{r})$. 

Consider first the effective couplings generated by in-medium two-pion and kaon exchange processes
and denoted by $G^\Lambda_{\pi,K}(\rho)$. Their contributions to the $\Lambda$ hyperon self-energy in the medium have been calculated explicitly \cite{KWhyp} as functions of the nuclear Fermi momentum. This calculation was performed using in-medium chiral SU(3) perturbation theory to two-loop order. Following the procedures described in Ref. \cite{Finelli0607} one finds:
\begin{equation}
G^\Lambda_{\pi,K}(\rho) = \Delta G^\Lambda + g_3\, \rho^{\frac{1}{3}} + 
g_4 \,\rho^{\frac{2}{3}}\;,~~~~~\Delta G^\Lambda = g_1 - g_2\,\bar{\Lambda}~. 
\label{Gchiral}
\end{equation}
The hyperon self-energy is derived and calculated in the non-relativistic limit at which Lorentz scalar and  vector contributions are indistinguishable. Our convention is that $G_{\pi,K}^\Lambda(\rho)$ acts with equal share in both scalar and vector channels so that its total contribution to the $\Lambda$ self-energy is $2G_{\pi,K}^\Lambda(\rho)\rho$ (ignoring the small difference between scalar and baryon densities at this point).
The density-independent piece $\Delta G^\Lambda$ (with constants $g_1$ and $g_2$) is associated with the regularization of divergent parts of two-pion exchange loop integrals, where $\bar{\Lambda}\simeq 0.7$ GeV is a typical cutoff scale. This high-momentum (or short-distance) piece is equivalent to a $\Lambda N$ contact term encoding unresolved short-range dynamics at momentum scales large compared to the nuclear Fermi momentum. The density dependent terms proportional to $g_3$ and $g_4$ reflect the action of the Pauli principle on intermediate nucleons
participating in the two-pion and kaon exchange processes. The constants $g_i$ are deduced from
 \cite{KWhyp}, with the following values:
 
  $~~~~~g_{1} =  2.51$ fm$^2$, $~~~~g_{2} = 0.83$ fm$^3$, $~~~~g_3 = -0.44$ fm$^3$, $~~~~g_{4} = 0.84$ fm$^4$.
  
Notably, in the terms with non-trivial density dependence representing in-medium chiral two-pion and kaon exchange dynamics, $g_3$ and $g_4$ enter with alternating signs. 

A second distinct set of contact terms, with coupling constants denoted as $G_V^{\Lambda(0)}$ and $G_S^{\Lambda(0)}$, is introduced to account for the strong Lorentz scalar and vector fields of about equal magnitude but with opposite signs, the ones that figure prominently in relativistic mean-field phenomenology \cite{SerotWalecka}. In the context of in-medium QCD sum rules \cite{inmediumqcdsumrules}, these terms can be associated with the leading density dependence of the chiral (quark) condensate, $\langle\bar{q}q\rangle$, and the quark density $\langle q^\dagger q\rangle$. 

Altogether, the scalar and vector $\Lambda$-nuclear interaction strengths
are given as:  
\begin{equation}
G^{\Lambda}_S({\rho}) = G^{\Lambda (0)}_S + G^\Lambda_{\pi,K}({\rho})~,
\quad \quad G^{\Lambda}_V({\rho}) = G^{\Lambda (0)}_V + G^\Lambda_{\pi,K}({\rho})~ .
\end{equation}
Note that with a typical cut-off parameter $\bar{\Lambda} \simeq 0.7$ GeV, the contact term from
chiral $2\pi$ exchange, $\Delta G^\Lambda \simeq -0.5$ fm$^2$, is
much smaller than what is expected for the individual magnitudes of the scalar and vector contact couplings $G_{S,V}^{\Lambda(0)}$. For nucleons, $G_{V}^{(0)}\simeq -G_{S}^{(0)} \sim 10$ fm$^2$. For the $\Lambda$ hyperon we also anticipate $G_{V}^{\Lambda(0)}\simeq -G_{S}^{\Lambda(0)}$, but with a typical strength of only about half that for nucleons. The approximate cancellation in the sum $\Sigma_S + \Sigma_V$ of the scalar and vector mean fields produced by $G_{S,V}^{(0)}$ is contrasted by their coherent enhancement in the difference 
$\Sigma_S-\Sigma_V$ that contributes prominently to the spin-orbit coupling. 

In summary, we have the low-density expansion
\begin{equation}
G^{\Lambda}_{S,V}(\rho) = \mathcal{G}^{\Lambda}_{S,V} + g_3\, \rho^{\frac{1}{3}} + 
g_4 \,\rho^{\frac{2}{3}}~,
\label{lowdenexp}
\end{equation}
with two contact terms, or low-energy constants
\begin{equation}
\mathcal{G}^{\Lambda}_{S,V} = G^{\Lambda(0)}_{S,V} + \Delta G^\Lambda~,
\label{contact}
\end{equation}
representing short-distance and ``vacuum" dynamics. Apart from the strength of the additional surface derivative term, these two constants are effectively
the only adjustable parameters of the model.  Their arrangement  
in the form (\ref{contact}) is such that individually large ``vacuum" parts 
$G^{\Lambda(0)}_{S,V}$ are separated from the smaller
piece $\Delta G^\Lambda$ arising from in-medium chiral perturbation theory in the non-relativistic
limit. Such a separation is not necessary in principle but useful in practice. In particular, it helps 
interpreting  the physics content of $G^{\Lambda(0)}_{S,V}$ in relation to in-medium QCD sum rules,
as will be discussed in Section \ref{QCDsumrules}.
  
Concerning the coupling strength $D^{\Lambda}_{S}$ of the surface (gradient) term in Eq. (\ref{E_lambda_int}) we follow a similar procedure as in Ref. \cite{Finelli0607}:  $D^{\Lambda}_{S}$ 
will first be treated as an adjustable parameter, and its resulting value will then be 
compared with an estimate based on in-medium chiral perturbation theory (cf. Appendix \ref{Sec1App}).
The coupling $D^{\Lambda}_{S}$ of the gradient term\footnote{
In non-relativistic calculations such gradient terms have usually been omitted on the grounds that 
finite-range effects are included in an approximate way by the use of the empirical charge densities \cite{MDG}, 
or they are thought to be absorbed in phenomenological non-linear terms in powers of  $\rho$ \cite{MDG,Skyrme,SKdef}.
} 
could be density dependent as well, but it will turn out in practical 
applications that it is not necessary to go beyond the simplest approximation 
with a constant coupling. 

\subsection{Single-particle equations}

The minimization of the hypernuclear ground-state energy leads to a set of 
coupled relativistic equations of Kohn-Sham type for 
the nucleons and the single $\Lambda$-hyperon (see Ref.~\cite{Finellihyper} for further details):
\begin{equation}
\label{dir_eq}
\left[ -i {\bm\gamma} \cdot {\bm\nabla} + M_i + \gamma_0\, \Sigma^i_V  + \Sigma^i_S \right] 
\psi^i_\alpha(\bm{r}) = \epsilon^i_\alpha \,\psi^i_\alpha(\bm{r})  \quad {\rm with} \; i=n, p, \Lambda \; ,
\end{equation}
where $\psi^i_\alpha(\bm{r})$ are the Dirac wave functions of single-particle orbitals $\alpha$ with energies $\epsilon_\alpha$ 
for the nucleons and the $\Lambda$ hyperon, $M_i$ are their corresponding 
masses, and  $\Sigma^i_{V,S}$  denote the vector and scalar 
self-energies, respectively. The hypernuclear ground state is determined 
by the self-consistent solution of the single-particle Kohn-Sham equations (\ref{dir_eq}) 
for a given number of protons, neutrons, and for the single $\Lambda$-hyperon. 
Self-consistency here means that the self-energies are functionals of 
the ground state density calculated in the {\it no-sea} approximation~\cite{SerotWalecka, FurnstahlLN} from the single-particle solutions 
of the Kohn-Sham equations for the nucleons. Because of the explicit density dependence 
of the couplings, rearrangement contributions \cite{Fuchs:1995as} appear in the 
vector self-energies of protons and neutrons. The $\Lambda$ self-energies 
\begin{equation}
\Sigma_V^\Lambda  = G^{\Lambda}_V(\rho)\, \rho~ ,~~~~~~~~~~~~~~~~
\Sigma_S^\Lambda  =  \left(G^{\Lambda}_S(\rho) + D^{\Lambda}_S\, \nabla^2\right) \rho_S  ,
\end{equation}
are then expressed in terms of the vector and scalar ground-state
local densities, $\rho(\bm{r})$ and $\rho_S(\bm{r})$, of the nuclear core. These densities are determined
self-consistently together with the wave functions  $\psi^{n,p}_\alpha(\bm{r})$ and  $\psi^\Lambda_\beta(\bm{r})$.

Note that the upper components of the $\Lambda$'s Dirac wave function in a
given orbital, $\psi^\Lambda_\beta$, experience a self-consistent potential ${\cal U}^\Lambda =  \Sigma_V^\Lambda + \Sigma_S^\Lambda$, the sum of the vector and scalar self-energies. Given that $G_{V}^{\Lambda(0)}$ and $G_{S}^{\Lambda(0)}$ almost cancel, we see that  ${\cal U}^\Lambda  \simeq 2\,G_{\pi,K}^\Lambda(\rho)\,\rho$  (+ surface~term) is of genuine non-relativistic origin. Moreover,
the dominant part of this average potential comes from the short distance (Hartree) piece proportional to $\Delta G^\Lambda$, while the sum of the $g_3$ and $g_4$ terms gives only a small correction.
The lower components of $\psi^\Lambda_\beta$, on the other hand, involve the large difference of vector and scalar self-energies, 
$\Sigma_V^\Lambda - \Sigma_S^\Lambda \simeq (G_{V}^{\Lambda(0)}-G_{S}^{\Lambda(0)})\,\rho$,
that enters in the discussion of the $\Lambda$-nuclear spin-orbit coupling, see subsection \ref{secth}. 

\subsection{\label{QCDsumrules} Guidance from in-medium QCD sum rules}  

The QCD ground state (vacuum) is characterized by condensates of 
quark-antiquark pairs and gluons, an entirely non-perturbative phenomenon. 
The quark condensate $\langle \bar{q}q \rangle$, i.e. the vacuum expectation 
value of the scalar quark density, plays a particularly important role 
as an order parameter of spontaneously broken chiral symmetry. 
At a renormalization scale of about 1 GeV the chiral vacuum 
condensate is $\langle \bar{q} q \rangle_0 \simeq - (240~
{\rm MeV})^3 \simeq - 1.8~{\rm fm}^{-3}$ \cite{Pic}. Hadrons, as well as nuclei, are 
excitations built on this condensed QCD ground state. The density-dependent changes of 
the condensate structure in the presence of baryonic matter are a source of 
strong scalar and vector mean fields experienced by nucleons (and hyperons).
In-medium QCD sum rules \cite{inmediumqcdsumrules} relate the leading changes 
of the scalar quark condensate $\langle \bar{q}q \rangle_\rho$ and quark density $\langle q^\dagger q \rangle = 3\rho$, at finite baryon density $\rho$, to the scalar and 
vector self-energies of a nucleon (or hyperon) in the nuclear medium.

The strength of the chiral condensate at normal nuclear matter density, $\rho_0 \simeq 0.16$ fm$^{-3}$, is reduced by 
about one third from its vacuum value. 
The detailed density dependence of this condensate has recently been studied \cite{KdeHW08} 
using in-medium chiral perturbation theory to three-loop order in the energy density, i.e.
at a level consistent with the approach employed in the present work. It is found that at densities
$\rho\lesssim\rho_0$, the leading linear $\rho$ dependence of $\langle\bar{q}q\rangle_\rho$ dominates
whereas non-linear effects become increasingly important at higher densities.
Assuming that the nucleon mass $M_N$ in vacuum scales roughly with the vacuum chiral condensate, 
the in-medium scalar and vector self-energies of the nucleon can be 
expressed as~\cite{inmediumqcdsumrules}:
\begin{equation}
\label{back1}
   \Sigma_S^{N(0)} =  
   - \frac{\sigma_N M_N}{m_\pi^2 f_\pi^2} \rho_S ~~,~~~~~  \Sigma_V^{N(0)} = \frac{4 (m_u + m_d)M_N}{m_\pi^2f_\pi^2} \rho ~~ .
\end{equation}
Here $\sigma_N = \langle N| m_q \,\bar{q} q |N \rangle$ is
the nucleon sigma term ($\simeq 50$ MeV), $m_\pi$ is the pion mass (138 MeV),
and $f_\pi = 92.4$ MeV is the pion decay constant. For the quark masses we take 
$m_u + m_d \simeq 12$ MeV (again at a renormalization scale of about 1 GeV).
The resulting $\Sigma_S^{N(0)}$ and $\Sigma_V^{N(0)}$ are 
individually large, $300 - 400$ MeV in magnitude. 
Their ratio $\Sigma_S^{N(0)}/\Sigma_V^{N(0)}\simeq -\sigma_N/4(m_u+m_d)\sim -1$
suggests the already mentioned strong cancellation of scalar and
vector potentials in the single-nucleon Dirac equation.
 
The constraints implied by Eq. (\ref{back1})
are admittedly not very accurate, given corrections 
from condensates of higher dimension and uncertainties in the values of
$\sigma_N$ and $m_u+m_d$. The estimated error for the ratio 
$\Sigma_S^{(0)}/\Sigma_V^{(0)}\simeq -1$ is about 20\%. Nonetheless,
Eq. (\ref{back1}) is useful for first orientation when estimating the contact couplings $G_{S,V}^{(0)}$:\begin{equation}
G_S^{(0)} \simeq  - \frac{\sigma_N M_N}{m_\pi^2 f_\pi^2} \simeq -{\sigma_N\over 4(m_u+m_d)}\,G_V^{(0)}\sim -11 \; {\rm fm}^2~~,
\end{equation}
using $\sigma_N\simeq 50$ MeV and $m_u + m_d \simeq 12$ MeV.
This estimate is actually in remarkable agreement with the values determined from 
a {\it best fit} analysis of ground-state properties of finite nuclei throughout the nuclear chart
($G_S^{(0)} =-11.5 $ fm$^2$ and $G_V^{(0)} = 11.0$ fm$^2$) \cite{Finelli0607}, the values we use as basic input in the present work as well.

In the case of a hyperon in the nuclear medium, finite-density QCD 
sum rules \cite {QCDsumrulehyp} predict reduced scalar and vector self-energies of 
the $\Lambda$ while maintaining $\Sigma_S^{\Lambda (0)} \simeq -\Sigma_V^{\Lambda (0)}$,
though with large uncertainties. We introduce the ansatz
\begin{equation}
G_{S,V}^{\Lambda (0)} = \zeta\, G_{S,V}^{(0)} ~~,
\label{zeta}
\end{equation}
with a parameter $\zeta< 1$ controlling the reduction of the $\Lambda$ couplings relative to those for the nucleon. Assuming that only non-strange quarks contribute to the contact 
interactions generated by the condensate background, a naive quark model estimate gives 
$\zeta = 2/3$.
We leave room for an optimization \cite{Finellihyper}\footnote{In Ref. \cite{Finellihyper} we have used $\chi$ instead of $\zeta$ for the ratios $G^{\Lambda(0)}/G^{(0)}$.} of the parameter $\zeta$ by comparison with empirical $\Lambda$ single-particle energies in hypernuclei. In fact, while the quark model value is an option, the result of the detailed QCD sum rule analysis \cite{QCDsumrulehyp} is closer to $\zeta = 0.4$. 

\section{Single particle states of the $\Lambda$}
Given the explicit input for the long and intermediate  range kaon and two-pion exchange interactions, the only remaining unknowns are the strengths of the short-distance (contact) terms, $G^{\Lambda(0)}_{S,V}$, and of the surface gradient term, $D_S^\Lambda$. These will now be fixed by detailed fits to the empirical single particle orbits of the $\Lambda$ in selected hypernuclei. The contact terms are specified by the parameter $\zeta$ relating the scalar and vector mean fields of the $\Lambda$ to those
of the nucleons, see Eq.(\ref{zeta}). The remaining $\Delta G^\Lambda$ involves the cutoff scale $\bar{\Lambda}$. One expects this cutoff scale to be around 0.7 GeV, subject to possible further fine-tuning.

We proceed as follows.  The values of  $\bar{\Lambda}$ and $D^{\Lambda}_S$ are 
adjusted for different values of $\zeta$ (using $0.4, 0.5$ and 
$2/3$, i.e. ranging from the QCD sum-rule estimate to the naive quark-model 
prediction), by performing a {\em least-squares} fit to the empirical 
single-$\Lambda$ energy levels:
\begin{equation}
\label{fit}
\chi^2 = \sum_\alpha\left( \frac{\epsilon^{th}_\alpha - \epsilon^{exp}_\alpha}
{\delta \epsilon^{exp}_\alpha}\right)^2 \; .
\end{equation}
Here $\epsilon^{th}_\alpha$ and $\epsilon^{exp}_\alpha$ are the theoretical 
and experimental single-$\Lambda$ energies, respectively, with 
uncertainties $\delta \epsilon^{exp}_\alpha$. 
For the set of experimental energies in Eq. (\ref{fit}) we choose the
{\it s-} and {\it p-} levels  of $^{16}_{~\Lambda} $O and  $^{208}_{\; \; \Lambda}$Pb
(see Table \ref{tab1})\footnote{
In a recent Jlab Hall-A report \cite{L16N}, Cusanno {\it et al.} studied
the single particle spectrum of $^{16}_{~\Lambda}$N and found
for the $1s$ binding energy the value $13.76 \pm 0.16$ MeV, 
in apparent disagreement with the corresponding value for $^{16}_{~\Lambda}$O
if charge symmetry would be approximately realized.
}.
An uncertainty of $3 \%$ is assumed for the
{\it s}-states, and $5\%$ for the {\it p}-states. 
For the nucleon sector the FKVW energy density functional \cite{Finelli0607} 
is used, without readjustments of any parameters. The fits are summarized 
in Table \ref{tab0}, where we display the optimal 
values of the cut-off scale $\bar{\Lambda}$ (or equivalently, the contact term proportional
to $\Delta G^\Lambda$), and the strength of the gradient term $D^S_\Lambda$,
for $\zeta= 0.4, 0.5$ and $2/3$. The corresponding $\chi^2$ values are 
given in the column on the right. Note that at this stage the data set used in the fit 
does not include the (small) splittings between the $p$-shell spin-orbit partner states, but 
only the average energies of the $p$-orbitals. 

\begin{table}[]
\caption{\label{tab0}Best-fit values of the contact term $\Delta G^\Lambda$ 
(determined by the cut-off scale $\bar{\Lambda}$) 
and of the strength $D^S_\Lambda$ of the surface (gradient) term,
for different choices of the ratio $\zeta = G_{S,V}^{\Lambda(0)}/G_{S,V}^{(0)}$ (input: $G_{S}^{(0)} = -11.5$ fm$^2$, $G_{V}^{(0)} = 11.0$ fm$^2$ from Ref.\cite{Finelli0607}). The fits are performed by reproducing single particle energies of $s$- and $p$-states in $^{16}_{~\Lambda}$O and $^{208}_{~~\Lambda}$Pb. Also shown is the resulting depth ${\cal U}^\Lambda$ of the $\Lambda$ central potential.}
\vspace{.3cm}
\begin{center}
{\small
\begin{tabular}{|ccccc||c|}
\hline
$\zeta$ & $\Delta G^\Lambda$ (fm$^2$) & $\bar{\Lambda}$  (MeV) & $D^S_\Lambda$ (fm$^4$) & 
${\cal U}^\Lambda$ (MeV) & $\chi^2$ \\
\hline \hline $0.4$ & $ -0.55 $ & $ 718.6$ & $-0.304 $ & $ -34.7 $ & $ 3.41$  \\
$0.5$ & $ -0.56 $ & $721.0 $ & $-0.340 $ & $ -35.4 $ & $4.34 $  \\
${2\over 3}$ & $ -0.58 $ & $726.1 $ & $-0.415 $ & $-36.6 $ & $2.86 $  \\
\hline 
\end{tabular}
}\end{center}
\vspace{0.5cm}
\end{table}

Quite acceptable fits are obtained with $\Delta G^\Lambda$ in the range $-(0.55 - 0.58)$ fm$^2$,
producing a central $\Lambda$ single particle potential ${\cal U}^\Lambda_{central} = 2\, \Delta G^\Lambda\rho(r) \simeq -(35 - 37)\,$MeV$\,(\rho({\bm r})/\rho_0)$. The required cutoff scales $\bar{\Lambda}$ are close to $\bar{\Lambda} \simeq 0.7$ GeV  \cite{KWhyp}, almost independent
of the ratio $\zeta = G^{\Lambda(0)}_{S,V}/G^{(0)}_{S,V}$. In view of the cancellation between scalar and vector contact terms $G^{(0)}_{S}$ and $G^{(0)}_{V}$, this is of course not surprising.

The inclusion of the surface term  proportional to ${{\bm \nabla}}
\rho_\Lambda \cdot {\bm \nabla} \rho$ turns out to be important. 
An attempt to fit the $\Lambda$ energy levels without this term results in an unacceptably large
$\chi^2$ value ($\simeq 200$), because it is not possible to 
simultaneously reproduce levels in light and heavy hypernuclei. 
For the strength of this term, $D^{\Lambda}_S$, one could directly use the 
theoretical estimate $-0.56$ fm$^4$ from in-medium chiral perturbation theory (cf. Appendix \ref{Sec1App}). 
The resulting agreement with data is then significantly improved, but not yet optimal ($\chi^2 \simeq 10$). In particular, one still encounters overbinding for heavy systems. The best fit values, 
$D^{\Lambda}_S \simeq -(0.3 - 0.4)$ fm$^4$,  are nevertheless still in remarkable qualitative agreement with chiral theory.

In this work all calculations are performed assuming spherical symmetry and a simplified configuration: a closed core of nucleon pairs plus a single 
$\Lambda$-hyperon. These are commonly used  approximations and recent investigations have
confirmed their validity. The study of Ref. \cite{SKdef} has shown 
that deformation effects need to be taken into account only in very light systems 
which are not the subject of the present investigation. Of the hypernuclei 
considered in this work, only $^{13}_{\; \Lambda}$C is 
sensitive to deformation, but  recent RMF studies \cite{hagino} have pointed out 
that the stabilizing effect of the $\Lambda$ tends to restore spherical symmetry 
in the hypernuclear system. Effects related to the odd number of nucleons in 
hypernuclei appear to be rather small \cite{SKdef}. In the case of hypernuclei 
with open nucleon shells, we include pairing correlations described in the BCS approximation 
with empirical pairing gaps \cite{MoellerNix} (except for $^{13}_{\; \Lambda}$C, 
which is calculated in the closed-shell approximation).

\section{\label{spinorbit}$\Lambda$-nuclear spin-orbit coupling}

Having specified the input for the $\Lambda$-hypernuclear central and surface
potentials, we can now concentrate on a more detailed investigation of the $\Lambda$-
nuclear spin-orbit interaction, one of the central themes of hypernuclear spectroscopy.
As already indicated in the introduction, $\Lambda$-hypernuclei feature 
extremely small energy spacings between spin-orbit 
partner states, as compared to the large spin-orbit splittings 
in ordinary nuclei \cite{hypspinorbit, hypspinorbit2, hypspinorbit3}. 
In this section we first briefly summarize the available data and
review the current theoretical approaches. The final subsection presents and 
discusses results based on our novel interpretation of 
the smallness of the $\Lambda$-nucleus spin-orbit 
coupling \cite{KWhyp, KWproposal}.  
 
\subsection{Brief summary of experimental results}


Detailed informations on $\Lambda$ spin-orbit splittings derive from measurements of light 
hypernuclei with nucleons in the {\it p}-shell. The BNL 929 
experiment \cite{929}, using $^{13}$C as target, reported an energy splitting between
the $p_{1/2}$ and $p_{3/2}$ $\Lambda$ levels 
$\Delta \epsilon^\Lambda (p) = 152 \pm 54 \pm 36$ keV. 
At present this experimental result provides the most convincing evidence 
for the small spin-orbit coupling in $\Lambda$-hypernuclei. It has been 
corroborated by experiments on heavier hypernuclei. 
For $^{16}_{\; \Lambda}$O a combined analysis of 
$(K^-,\pi^- \gamma)$ and $(\pi^+,K^+)$ experimental spectra 
(experiments KEK E336 \cite{336} and CERN-SPSII \cite{CERNO16}),  
allowed the determination of the very small energy level splitting between 
the $2^+$ and $0^+$ states: $\Delta \epsilon (2^+-0^+) = 40 \pm 320$ keV.  
Motoba {\it et al.} \cite{Motoba016} used the linear dependence of the relation
between $\Delta \epsilon(2^+-0^+)$ and the {\it p}-level splitting
$\Delta \epsilon^\Lambda(p)$, to estimate $\Delta \epsilon^\Lambda(p) = 300-600$ keV\footnote{ 
A more recent study by Hashimoto {\it et al.},  
lowers this estimate to: $-800 < \Delta \epsilon^\Lambda(p) < 200$ keV 
(cf. Fig. 23 in Ref. \cite{hypspinorbit}).}. 
An updated analysis \cite{Motoba_08} of data from the experiment 
$^{89}$Y($\pi^+,K^+$)$^{89}_{\; \Lambda}$Y \cite{y89exp} has
confirmed that the spin-orbit interaction is also
strongly suppressed in heavy hypernuclei: $\Delta \epsilon^\Lambda (f) \simeq 200$ keV, 
$\Delta \epsilon^\Lambda (d) \simeq150$ keV and $\Delta \epsilon^\Lambda (p) \simeq 90$ keV. 
So far these are the only empirical spin-orbit splittings outside the 
region of {\it p}-shell nuclei. 

\subsection{\label{secth}Previous theoretical studies}

The framework of energy density functionals provides an
accurate description of hypernuclei over the whole mass table \cite{MDG}.
At the same time light systems (e.g. {\it p}-shell hypernuclei) are also successfully  
decribed using phenomenological 
hyperon-nucleus potentials such as the one introduced by Dalitz and collaborators \cite{DalitzGal}. 
More recent {\it ab-initio} calculations start from 
realistic free $YN$ potentials, as for example the Nijmegen studies based on meson-exchange interactions \cite{nijmegen,esc04}, or the J\"ulich potential based on $SU(3)$ 
chiral perturbation theory \cite{julich}. However,  all these realistic $YN$ potentials 
tend to strongly overestimate the $\Lambda N$ spin-orbit interaction in hypernuclei.
For instance, recent calculations based on the cluster model approach \cite{spoth} 
predict $\Delta \epsilon^\Lambda(p) \simeq 390 - 960$ keV for 
$^{13}_{\; \Lambda}$C\footnote{
With a recent improvement of the Nijmegen potential, called ESC06 \cite{esc06},
it appears possible to obtain values of $\Delta \epsilon^\Lambda$ that are closer to data.
However, Nijmegen potentials such as ESC04, fail to 
reproduce the single-$\Lambda$ binding energies \cite{halderson}. Thus the microscopic interpretation of the $\Lambda$-nuclear spin-orbit interaction, starting from one-boson exchange $YN$ 
interactions, remains an open problem. For the latest version (ESC07) including also
contributions from quark degrees of freedom, see \cite{SENDAI}.}.

For first orientation, recall the  Walecka-type relativistic mean-field approach to nuclei \cite{SerotWalecka} in which short-range dynamics is parametrized by the exchange of a phenomenological scalar boson ($\sigma$) 
and a vector boson ($\omega$) between nucleons. 
The corresponding mean fields produce large scalar $(S)$ and vector $(V)$ nucleon self-energies. 
The effective spin-orbit potential \cite{BM} is obtained 
in the non-relativistic limit of the single-particle Dirac equation:
\begin{equation}
\label{so1}
  V_{so} = \frac{1}{2M^2} \left(
  {1 \over r}{\partial \over \partial r} V_{ls}(r)\right) {\bm l}\cdot {\bm s} \; ,
\end{equation}
where the large spin-orbit potential $V_{ls}$ arises
from the difference of the vector
and scalar potentials:  $\Sigma_V(r) = g_\omega \,\omega_0(r)$ and  $\Sigma_S(r) = g_\sigma \,\sigma(r)$, with $\Sigma_V(0)\simeq   330$ MeV and 
$\Sigma_S(0) \simeq - 400$ MeV \cite{BW77,ring96,vretenar05}. 
Explicitly,
\begin{equation}
\label{so2}
   V_{ls} = {M \over M_{eff}} (\Sigma_V - \Sigma_S) \; ,
\end{equation}
where
$M_{eff}$ is an effective mass specified as \cite{ring96}
\begin{equation}
\label{so3}
   M_{eff} = M - {1 \over 2} (\Sigma_V - \Sigma_S) \;.
\end{equation}

At this point it is useful to clarify the close correspondence between a Walecka type
phenomenology (or an in-medium QCD sum rule approach) of strong scalar-vector mean fields, and certain contact terms appearing at next-to-leading order (NLO) in the effective field theory description of the NN interaction.  A combination of NLO contact terms (with two derivatives, see Eq. (5) in Ref. \cite{Kaiser:2004sh}) generates a Galilei invariant spin-orbit interaction 
\begin{equation}
{-i\over 4}C_5\, ({\bm \sigma}_1 + {\bm \sigma}_2) \cdot \left[({\bm p}_1^{\;'}-{\bm p}_2^{\;'})  \times ({\bm p}_1-{\bm p}_2)\right] \;,
\end{equation}
where $C_5$ is given by a linear combination of $P$-wave low-energy constants:
\begin{equation}
C_5 = \frac{1}{16\pi} [ 2C(^3P_0) + 3C(^3P_1) -5 C(^3P_2)] \;.
\label{spinorbitcontact2}
\end{equation}
It has been shown in Ref. \cite{Kaiser:2004sh} that the strength of the spin-orbit term derived from the free $NN$ interaction agrees quantitatively with the one required in nuclear shell-model calculations.
It has furthermore been demonstrated in Ref. \cite{Plohl:2006hy} that the spin-orbit contact term 
(\ref{spinorbitcontact2}), in a relativistic Dirac-Br\"uckner (DBHF) calculation, generates the strong scalar and vector mean fields in the combination $\Sigma_S - \Sigma_V$. 
Thus the inclusion of contact terms representing condensate background fields 
does not at all spoil the consistency of in-medium chiral calculations based on the LO $\pi N$ Lagrangian. These background fields are just equivalent reflections of short distance $NN$ dynamics. 

The scalar-vector approach can be transcribed analogously for hypernuclei. Empirical spin-orbit splittings are reproduced by simply assuming {\it much weaker} couplings between 
the $\Lambda$ and the exchanged bosons. In particular, in one of the early studies 
of this type \cite{BrockmannWeise}, a reduction of 
$1/3$ was suggested for the $\Lambda$
potentials with respect to the corresponding nucleon self-energies
\begin{equation}
\Sigma_S^\Lambda = \frac{1}{3} \Sigma_S \quad {\rm and} \quad \Sigma_V^\Lambda = \frac{1}{3} \Sigma_V \; .
\end{equation} 
In contrast, the naive quark model assumes that the non-strange 
quarks couple to the $\sigma$ and $\omega$ mean-fields 
(the $s$-quark spectator hypothesis) and suggests a reduction factor of
$2/3$. With this value, however, it is not possible to reproduce the 
empirical spin-orbit splittings in $\Lambda$-hypernuclei. A possible 
solution proposed in Ref. \cite{noblepirner} involved an 
additional strong tensor coupling term in the $\omega \Lambda$
interaction Lagrangian
\begin{equation}
\mathcal{L}_{\omega \Lambda} = 
g_\omega^\Lambda \bar{\psi}_\Lambda \gamma^\mu \psi_\Lambda \omega_\mu +
\frac{f_\omega^\Lambda}{2M_\Lambda} \bar{\psi}_\Lambda
\sigma^{\mu \nu} \psi_\Lambda \partial_\nu \omega_\mu \;.  
\end{equation}
This additional term modifies the effective $\Lambda$ spin-orbit potential as follows:
 \begin{equation}
 \label{so_phen}
V_{so,\Lambda} \simeq \frac{1}{2M^{*2}_\Lambda} \left[\frac{1}{r} \frac{\partial}{\partial r} \left(
\left(2\frac{f_\omega^\Lambda}{g_\omega^\Lambda} +1 \right) \Sigma_V^\Lambda-\Sigma_S^\Lambda \right) \right] 
{\bm l} \cdot {\bm s}\; .
\end{equation}
For $f_\omega^\Lambda / g_\omega^\Lambda = -1$ the potential 
$V_{ls}(\Lambda) = (2{f_\omega^\Lambda}/{g_\omega^\Lambda} +1 )
\Sigma_V^\Lambda-\Sigma_S^\Lambda $ is now very small compared to that for the nucleon 
(see Ref. \cite{jennings} for more details, and Figs. 2 and 3 of Ref. \cite{tensor}). 

While phenomenological studies, based on the assumption of a 
strong $\omega \Lambda$ tensor coupling \cite{tensor}, worked in reproducing 
the empirical single-$\Lambda$ levels for a number of hypernuclei, 
they did not offer a consistent microscopic 
explanation for the spin-orbit suppression in $\Lambda$-hypernuclei.
Realistic $YN$ potentials (NSC97) that reproduce phase shift data suggest significantly weaker 
$\omega \Lambda$ tensor couplings \cite{nijmegen}. The older
Nijmegen D and F potentials, for example, give -0.12 and -0.54
for the ratio $f_\omega^\Lambda/g_\omega^\Lambda$ \cite{Gal_insertion}, respectively, considerably
smaller in magnitude than the value $\approx -1$ required in 
Eq.(\ref{so_phen}) to reproduce the empirical spin-orbit splittings.

Alternative microscopic models of hypernuclear spectroscopy 
have also been developed more recently. A synthesis of quark-model 
and relativistic one-boson exchange picture has been established 
by the Quark-Meson Coupling (QMC) model~\cite{Tsushima_old,Tsushima_new}. In this model 
$V_{so,\Lambda}$ arises entirely from Thomas precession. To obtain the correct spin-orbit splittings,
a piece
$- \frac{2}{M_\Lambda^{*2}r}g_\omega^\Lambda\frac{d}{dr} \omega(r)\,{\bm l} \cdot {\bm s}$
must be included in addition to the self-consistent calculation of single-$\Lambda$ energy levels.
Lenske {\it et al.} ~\cite{Keil:1999hk} have 
developed a density-dependent relativistic framework
in which the $\Lambda$-meson couplings are partly determined 
from a $\Lambda$N T-matrix, and partly fitted to a selected set of data.
In that approach the spin-orbit energy splittings $\Delta \epsilon^\Lambda$ 
display a uniform dependence on the nuclear mass number $A$
(cf. Fig. 7 of Ref. \cite{Keil:1999hk}). This is in contrast 
to the QMC results, where $\Delta \epsilon^\Lambda \simeq 0$ for all hypernuclei and 
any $\Lambda$ orbital \cite{Tsushima_old}. 
The spin-orbit energy splittings computed in \cite{Keil:1999hk} still overestimate the 
empirical values, and this led those authors to raise 
the quest for an additional reduction mechanism of the $\Lambda$-nuclear
spin-orbit force.

 \subsection{\label{chiralspinorbit}The $\Lambda$-nuclear spin-orbit interaction
 from chiral SU(3) two-pion exchange} 
 
 Starting from in-medium chiral $SU(3)$ dynamics for the $\Lambda$ hyperon in nuclear
 matter \cite{KWhyp, KWproposal} and its translation to a hypernuclear energy density
 functional \cite{Finellihyper}, we have identified two basic competing mechanisms at the 
 origin of the unusually small $\Lambda$-nuclear spin-orbit interaction: strong scalar-vector mean
 fields acting coherently, and the spin-orbit force of opposite sign \cite{KWhyp}
 induced by the second order pion exchange tensor interaction with an intermediate $\Sigma$
 hyperon. A third prominent contributor to the spin-orbit force in ordinary nuclei, namely the
 three-body interaction of Fujita-Miyazawa type, has no counterpart in single-$\Lambda$ 
 hypernuclei, as already mentioned.
 
 Let us briefly recall the steps leading to the ``wrong sign" spin-orbit interaction from two-pion
 exchange \cite{KWhyp}. Consider a $\Lambda$ that scatters in slightly inhomogeneous nuclear matter 
from an initial momentum ${\bm p} - {\bm q}/2$ to a final momentum ${\bm p} + {\bm q}/2$. One identifies a spin-orbit term 
$\Sigma_{ls}^\Lambda(k_F) = 
{i\over 2}U_{ls}^\Lambda(k_F)\,
{\bm\sigma}\cdot({\bm q}\times {\bm p}\,)$, 
in the spin-dependent self-energy of the $\Lambda$, with
\begin{eqnarray}
U_{ls}^\Lambda(k_F) = & &-{2\over 3}\left({D\,g_A\over f_\pi^2}\right)^2\int_{|{\bm p}_1|<k_F}{d^3p_1\over (2\pi)^3}\int_{|{\bm p}_2|>k_F}{d^3p_2\over (2\pi)^3} \nonumber\\
& &\hspace{3cm} {({\bm p}_1 - {\bm p}_2)^4 M_B\over [m_\pi^2 + ({\bm p}_1 - {\bm p}_2)^2]^2\,[\Delta^2 + {\bm p}_2\,^2 - {\bm p}_1\cdot {\bm p}_2]^2}~.
\label{spinorbitloop}
\end{eqnarray}

\begin{figure}[t]
\centering
\includegraphics[scale=0.35,angle=0]{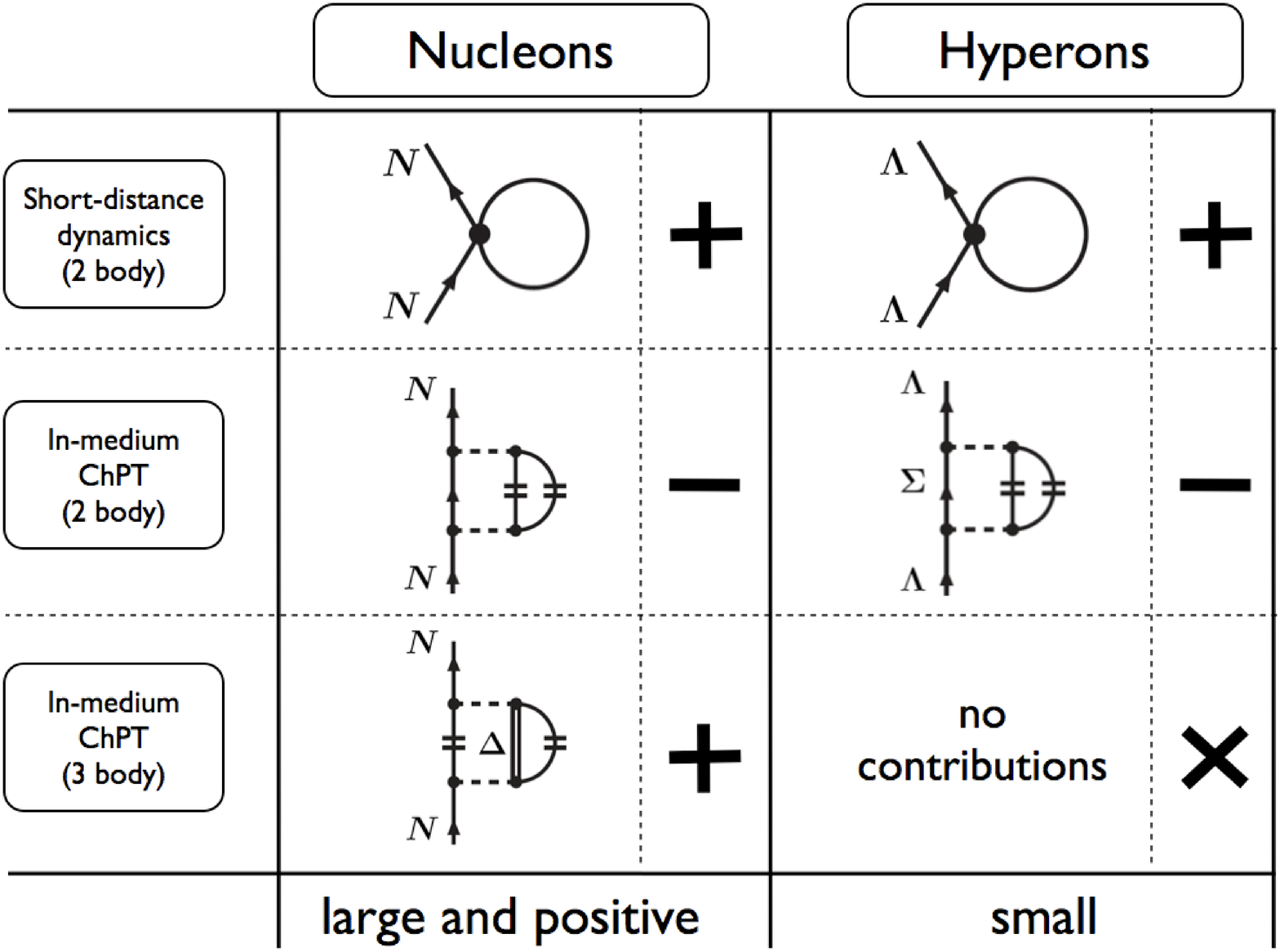}
\caption{\label{figA} Schematic picture of contributions to the spin-orbit interaction of a nucleon in nuclei and  a $\Lambda$ in hypernuclei. First row: contact terms summarizing strong scalar and vector fields
and unresolved short-distance dynamics. Second row: iterated pion exchange (mostly second order 
tensor interactions) with Pauli-blocked intermediate nucleon states.
Third row: three body terms generated by $2\pi$ exchange with 
intermediate excitation of a virtual $\Delta$ isobar and in-medium insertions (resembling
Fujita and Mijazawa \cite{fujita}). 
This mechanism has no counterpart for $\Lambda$ hyperons since there is no filled hyperon Fermi sea.    
Signs of the different contributions are indicated.}
\vspace{0.3cm}
\end{figure}

It depends only on known $SU(3)$ axial-vector 
coupling constants ($D$ = 0.84, $g_A$ = 1.3) and on $\Delta^2 = (M_\Sigma - M_\Lambda)M_B$ which involves the (small) mass difference  
between the $\Sigma$ and the $\Lambda$ hyperon. The average baryon mass $M_B =$ 1.05 GeV appearing in the numerator is a reminder that this spin-orbit interaction has a {\it non-relativistic} origin. The momentum space loop integral (\ref{spinorbitloop}) is finite and hence model independent, 
in the sense that no regularizing cutoff is required. 
The spin-orbit coupling strength at saturation density, 
$U_{ls}^\Lambda(k_F^{(0)})\simeq -15.1$ MeV fm$^2$ 
at $k_F^{(0)} \simeq 1.36$ fm$^{-1}$, has a sign {\it opposite}
to the standard spin-orbit coupling strength. 
Evidently, this contribution to the $\Lambda$ spin-orbit 
potential tends to cancel the contribution 
from the strong scalar-vector mean fields. 
However, contrary to the case of nucleons in ordinary nuclei (see Fig. \ref{figA}),
this ``wrong sign" spin-orbit interaction is not 
compensated in turn by a three-body spin-orbit interaction \cite{KWhyp,KWproposal}. 
Thus the smallness of the $\Lambda$-nucleus spin-orbit finds its natural 
explanation in terms of an almost complete cancellation between 
short-range background mean-field contributions and longer range 
terms generated by  $2\pi$-exchange.


One might argue that the $U_{ls}^\Lambda$ of Eq. (\ref{spinorbitloop}), derived for infinite nuclear matter with its continuous
single particle energy spectrum, overestimates the magnitude of the two-pion exchange spin-orbit force considerably. This turns out not to be the case.
In order to examine this question, consider instead a finite cubic box of length $2L$ with a discrete
momentum spectrum, ${\bm p} = (\pi/L){\bm n}$ and ${\bm n}\in {\cal Z}^3$. Momentum space integrals are now replaced by sums, $\int d^3p/(2\pi)^3 \rightarrow (1/8L^3)\sum_{{\bm n}}$. The density $\rho$ is $1/(2L^3)$ times the sum over all occupied, discrete single particle states. Now take, for example, a finite system with $A = 48$ for which $L\simeq 3.35$ fm. The result is $U_{ls}^\Lambda \simeq -14.7$ MeV fm$^2$, to be compared with -15.1 MeV fm$^2$ for nuclear matter. 

The quantitative effect of in-medium two-pion exchange on the energy 
spacings between single-$\Lambda$ spin-orbit partner states, is calculated 
in first-order perturbation theory:
\begin{equation} 
\label{so}
\Delta\epsilon^{\Lambda}_{\alpha} =
\langle\alpha | \Delta\mathcal{H}_{ls}^{\Lambda} | \alpha \rangle \;,
\label{eqn:ls}\end{equation}
for each hyperon orbit $\alpha$, with 
\begin{equation}
\Delta\mathcal{H}_{ls}^{\Lambda} = 
{U_{ls}^\Lambda(k_F^{(0)})\over 2r}\, \frac{d}{dr} \left(\frac{\rho(r)}{\rho^{(0)}} \right) \,
{\bm\sigma}\cdot {\bm l} \; .
\end{equation}
The $\Lambda$ single particle states $|\alpha\rangle$ are determined by 
self-consistent solutions of the Dirac  
equation (\ref{dir_eq}), and $\rho(r)$ is the corresponding ground-state nuclear density, also
computed self-consistently, with $ \rho^{(0)}  \equiv \rho(r=0)$.


\begin{figure}[h]
\begin{center}
\includegraphics[scale=0.35,angle=0]{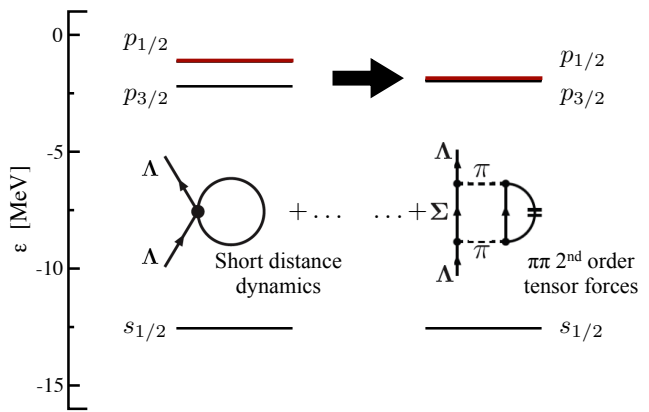}
\caption{\label{figBbis} Calculated $\Lambda$ single particle energy levels in  $^{16}_{\; \Lambda}$O.
Left panel shows the spin-orbit splitting from short-distance dynamics (scalar and vector mean fields)
only; right panel: after inclusion of the spin-orbit term from in-medium two-pion exchange.}
\end{center}
\vspace{.3cm}
\end{figure}


A demonstration of the counterbalance between short-distance contact term and intermediate range two-pion exchange contributions to the spin-orbit splitting is shown in Fig. \ref{figBbis} for the example of 
$^{16}_{\; \Lambda}$O. It shows how the spin-orbit term from the in-medium two-pion exchange (second-order tensor) force basically compensates the contribution from scalar-vector mean fields.

We close this subsection with a note on corrections to the chiral spin-orbit potential from the $SU(3)$ decuplet. A study along these lines has been performed in Ref. \cite{camalich} which extended the work of Ref. \cite{KWhyp} and considered also $\Delta$ and $\Sigma^*$ intermediate states in the two-pion exchange processes. Corrections from virtual excitations  of such states, despite their large masses,
turned out to be non-negligible because of their strong couplings to the pions and the baryons octet.
The results found previously in Ref. \cite{KWhyp} were basically confirmed:
the spin-orbit potential from two-pion exchange is a non-relativistic effect as demonstrated by its proportionality to the baryon mass. It is model independent as it does not require regularization and depends only on physical parameters (masses and coupling constants). We have examined this extended $U^\Lambda_{ls}$ in our approach as well and briefly report results at the end of the next section. 

\section{Results}

The previously decribed framework, with parameters fixed once and for all
by the $s$- and $p$-states of  $^{16}_{\; \Lambda}$O and $^{208}_{\; \;\Lambda}$Pb, is now
tested systematically in comparison with the empirical energies for all known single-$\Lambda$ levels of six hypernuclei, from light to heavy. The input from Table \ref{tab0} with $\zeta =0.5$ is used\footnote{We recall that the results are not sensitive to a particular choice of parameter sets in Table \ref{tab0}.}, together with the FKVW parameters \cite{Finelli0607} in the nucleon sector of the energy density functional but, as will shortly be shown,
the results are not very sensitive to a particular choice of
one of the three parameter sets for the $\Lambda N$ couplings.

Our results are summarized in Tables \ref{tab1} and \ref{tab2} and compared with those of six different calculations: 
Quark Meson Coupling (QMC) \cite{Tsushima_old, Tsushima_new}, Fermi Hypernetted Chain (FHNC) \cite{FHNC}, Skyrme (SK) \cite{Skyrme}, Brueckner-Hartree-Fock (BHF) \cite{BHF} with the Njimegen SC97F potential \cite{nijmegen}, relativistic mean field models with a tensor coupling \cite{tensor} (RMFI with $f_\omega^\Lambda/g_\omega^\Lambda = -1$) and density-dependent interactions \cite{Keil:1999hk} (RMFII). 
The single particle energy spectra are calculated in the approximation of a closed 
even-even nucleon core + $\Lambda$-hyperon, i.e. the theoretical 
spectra\footnote{for $^{13}_{\,\Lambda} $C the calculation should not be considered fully realistic due to
the closed shell approximation.
} 
correspond to $^{13}_{\,\Lambda} $C, $^{17}_{\,\Lambda}$O, 
$^{41}_{\,\Lambda}$Ca, $^{91}_{\,\Lambda} $Zr, $^{141}_{~\Lambda} $Ce and 
$^{209}_{~\Lambda} $Pb. 

Comparing our (FKVW) results to those of a relativistic mean-field
calculation (RMF1) that uses a strong $\Lambda$-nuclear tensor force
with $f_\omega^\Lambda/g_\omega^\Lambda = -1$, one observes differences
in the spin-orbit splittings of $\Lambda$ orbitals with large angular
momentum $l$. For example, the FKVW spin-orbit splitting, though
very small on an absolute scale and well within experimental errors,
tends to be more than twice the RMF1 splitting for a $\Lambda$ in $f$-
and $g$-orbitals. The mechanisms at work in FKVW and RMF1 involve of
course very different physics. A technical difference arises because
the FKVW spin-orbit splitting is treated perturbatively, whereas the
RMF calculations are performed self-consistently.


\begin{table}[h]
\caption{\label{tab1} Binding energies (in MeV) of single-$\Lambda$ levels in $^{13}_{~\Lambda}$C, 
$^{16}_{~\Lambda} $O, $^{40}_{~\Lambda }$Ca and $^{89}_{~\Lambda }$Y.
Experimental energies \cite{hypspinorbit} are shown in comparison with the
results of the present calculations, using the input parameters of Table \ref{tab0} and $\zeta= 0.5$  (column FKVW). Also listed are results of five different
models: Quark Meson Coupling (QMC) \cite{Tsushima_old,Tsushima_new}, Fermi Hypernetted Chain (FHNC) \cite{FHNC}, Skyrme (SK) \cite{Skyrme}, Brueckner-Hartree-Fock (BHF) \cite{BHF} with the Njimegen SC97F potential \cite{nijmegen}, and RMF models with a tensor coupling \cite{tensor} (RMFI with $f_\omega^\Lambda/g_\omega^\Lambda = -1$) and density-dependent couplings \cite{Keil:1999hk} (RMFII). 
}
\vspace{.3cm}
\begin{center}
{\small
\begin{tabular}{|ccc||c||cccccc|}
\hline
Nucleus & $\epsilon_{s.p.}$ & Expt. & FKVW & QMC & FHNC  & SK & BHF & RMFI & RMFII\\
\hline \hline $^{13}_{~\Lambda}$C & $1s_{1/2}$ & $ 11.38\pm 0.05 $ & $ 12.3 $ & $ -$ & $ 8.3 $ & $ 11.7 $ & $ 13.7 $ & $ 12.5 $ & $ 11.7 $\\
\hline $\;$ & $\begin{array}{c} 1p_{3/2} \\ 1p_{1/2} \end{array}$ & $ 0.38 \pm 0.1 $ & $\begin{array}{c} 0.1 \\ 0.0 \end{array} $ & $- $ & $-$ & $0.9 $ & $ 1.4$ & $\begin{array}{c} 1.1 \\ 0.8 \end{array} $ & $\begin{array}{c} 1.1 \\ 0.0 \end{array} $\\
\hline \hline
$^{16}_{~\Lambda} $O & $1s_{1/2}$ & $ 12.42 \pm 0.05 $ & $ 12.6 $ & $ 16.2 $ & $ 12.00 $ & $ 13.3 $ & $ 15.5 $ & 
$ 12.9$ & $ 12.8 $ \\
\hline $\;$ & $\begin{array}{c} 1p_{3/2} \\ 1p_{1/2} \end{array}$ & $ 1.85 \pm 0.06$ & $\begin{array}{c} 2.0 \\ 1.9 \end{array} $ & $\begin{array}{c} 6.4 \\ 6.4 \end{array} $ & $1.8 $ & $3.0 $ & $ 3.7$ & $ \begin{array}{c} 3.3\\ 3.0 \end{array} $ & $ \begin{array}{c} 2.8\\ 1.4 \end{array} $  \\
\hline \hline
$^{40}_{~\Lambda}$Ca & $1s_{1/2}$ & $ 20.0 \pm 1.0 $ & $ 18.9 $ & $ 20.6 $ & $ 20.0 $ & $ 18.0 $ & $ 20.7$ & $ 19.0$ & $ 17.6 $ \\
\hline $\;$ & $\begin{array}{c} 1p_{3/2} \\ 1p_{1/2} \end{array}$ & $ 12.0 \pm 1.0 $ & $ \begin{array}{c} 10.1 \\ 10.1 \end{array} $ & $\begin{array}{c} 13.9 \\ 13.9 \end{array} $ & $10.6 $ & $10.1 $ & $ 11.5$ & $ \begin{array}{c} 10.7 \\ 10.5 \end{array} $ & $ \begin{array}{c} 9.1 \\ 7.8 \end{array} $ \\
\hline $\;$ & $\begin{array}{c} 1d_{5/2} \\ 1d_{3/2} \end{array}$ & $ 1.0\pm 1.0 $ & $ \begin{array}{c} 1.6 \\ 0.9 \end{array} $ & $\begin{array}{c} 5.5 \\ 5.5 \end{array} $ & $1.6 $ & $1.6 $ & $ 2.0$ & $ \begin{array}{c} 2.7 \\ 2.4 \end{array} $ & $ \begin{array}{c} 1.5 \\ 1.5 \end{array} $ \\
\hline \hline
$^{89}_{~\Lambda} $Y & $1s_{1/2}$ & $ 23.1 \pm 0.5 $ & $ 23.4 $ & $ 24.0 $ & $ 23.3 $ & $ 21.1$ & $ 24.1 $ & 
$ 23.7$ & $ 23.2 $ \\
\hline $\;$ & $\begin{array}{c} 1p_{3/2} \\ 1p_{1/2} \end{array}$ & $ 16.5\pm 4.1 $ & $ \begin{array}{c} 17.2 \\ 17.2 \end{array} $ & $\begin{array}{c} 19.4 \\ 19.4 \end{array} $ & $16.9 $ & $15.6 $ & $ 17.8$ & $ \begin{array}{c} 17.6 \\ 17.4 \end{array} $ & $ \begin{array}{c} 17.2 \\ 16.3 \end{array} $ \\
\hline $\;$ & $\begin{array}{c} 1d_{5/2} \\ 1d_{3/2} \end{array}$ & $ 9.1\pm 1.3 $ & $ \begin{array}{c} 10.2 \\ 9.8 \end{array} $ & $\begin{array}{c} 13.4 \\ 13.4 \end{array} $ & $10.1 $ & $9.1$ & $ 10.4$ & $ \begin{array}{c} 10.7 \\ 10.5 \end{array} $ & $ \begin{array}{c} 10.3 \\ 8.9 \end{array} $ \\
\hline $\;$ & $\begin{array}{c} 1f_{7/2} \\ 1f_{5/2} \end{array}$ & $ 2.3 \pm 1.2 $ & $ \begin{array}{c} 2.8 \\ 2.0 \end{array} $ & $\begin{array}{c} 6.5 \\ 6.4 \end{array} $ & $- $ & $2.1$ & $ 2.4 $ & $ \begin{array}{c} 3.7 \\ 3.4 \end{array}$ & $ \begin{array}{c} 3.1 \\ 1.0 \end{array}$ \\
\hline 
\end{tabular}
}\end{center}
\end{table}
\newpage


\begin{table}[h]
\caption{\label{tab2} Binding energies (in MeV) of single-$\Lambda$ levels 
in $^{139}_{\;\; \Lambda} $La and $^{208}_{\;\; \Lambda} $Pb (continued from Table \ref{tab1}).}
\vspace{.3cm}
\begin{center}
{\small
\begin{tabular}{|ccc||c||cccccc|}
\hline
Nucleus & $\epsilon_{s.p.}$ & Expt. & FKVW & QMC & FHNC  & SK & BHF & RMFI & RMFII\\
\hline \hline
$^{139}_{\;\; \Lambda}$La & $1s_{1/2}$ & $ 24.5 \pm1.2 $ & $ 24.7 $ & $- $ & $ - $ & $ 22.1 $  & $ 25.3$ & $ 25.2$ & $ 25.2 $ \\
\hline$\;$ & $\begin{array}{c} 1p_{3/2} \\ 1p_{1/2} \end{array}$ & $ 20.4 \pm 0.6 $ & $ \begin{array}{c} 20.0 \\ 20.0 \end{array}  $ & $ -$ & $- $ & $17.9 $ & $ 20.5$  & $ \begin{array}{c} 20.4 \\ 20.4 \end{array}$ & $ \begin{array}{c} 20.5\\ 20.2 \end{array}$\\
\hline$\;$ & $\begin{array}{c} 1d_{5/2} \\ 1d_{3/2} \end{array}$ & $ 14.3 \pm 0.6 $ & $ \begin{array}{c} 14.3 \\ 14.1 \end{array} $ & $-$ & $-$ & $12.8 $ & $ 14.5$ & $ \begin{array}{c} 14.8 \\ 14.6 \end{array} $
 & $ \begin{array}{c} 14.9 \\ 14.1 \end{array} $\\
\hline $\;$ & $\begin{array}{c} 1f_{7/2} \\ 1f_{5/2} \end{array}$ & $ 8.0 \pm 0.6 $ & $ \begin{array}{c} 8.0 \\ 7.4 \end{array} $ & $- $ & $- $ & $6.9$ & $ 7.8$ & $ \begin{array}{c} 8.6 \\ 8.4 \end{array}$  & $ \begin{array}{c} 8.5 \\ 7.1 \end{array}$\\
\hline $\;$ & $\begin{array}{c} 1g_{9/2} \\ 1g_{7/2} \end{array}$ & $ 1.5 \pm 0.6 $ & $ \begin{array}{c} 1.5 \\ 0.5 \end{array} $ & $- $ & $- $ & $0.6 $ & $ 0.6$ & $ \begin{array}{c} 2.4 \\ 2.0 \end{array}  $ & $ \begin{array}{c} 2.2 \\ 0.2 \end{array}  $\\
\hline
\hline
$^{208}_{\;\; \Lambda}$Pb & $1s_{1/2}$ & $ 26.3 \pm 0.8 $ & $ 25.8 $ & $ 26.9 $ & $ 27.6 $ & $ 23.1 $ & $26.5 $  & $ 26.5$ & $ 27.2 $ \\
\hline $\;$ & $\begin{array}{c} 1p_{3/2} \\ 1p_{1/2} \end{array}$ & $ 21.9 \pm 0.6 $ & $ \begin{array}{c} 22.0 \\ 22.0 \end{array} $ & $ \begin{array}{c} 24.0 \\ 24.0 \end{array} $ & $22.8 $ & $19.6 $ & $ 22.4$ & $ \begin{array}{c} 22.7 \\ 22.6 \end{array} $ & $ \begin{array}{c} 23.4 \\ 23.1 \end{array} $\\
\hline $\;$ & $\begin{array}{c} 1d_{5/2} \\ 1d_{3/2} \end{array}$ & $ 16.8\pm 0.7 $ & $ \begin{array}{c} 17.4 \\ 17.3 \end{array} $ & $ \begin{array}{c} 20.1 \\ 20.1 \end{array} $ & $17.4 $ & $15.4 $ & $ 17.5$ & $ \begin{array}{c} 18.0 \\ 17.9 \end{array} $ & $ \begin{array}{c} 18.5 \\ 17.9 \end{array} $\\
\hline $\;$ & $\begin{array}{c} 1f_{7/2} \\ 1f_{5/2} \end{array}$ & $ 11.7 \pm 0.6 $ & $ \begin{array}{c} 12.2 \\ 11.8 \end{array} $ & $ \begin{array}{c} 15.4 \\ 15.4 \end{array} $ & $- $ & $10.5 $ & $ 11.8$ & $ \begin{array}{c} 12.7 \\ 12.5 \end{array} $ & $ \begin{array}{c} 13.2 \\ 12.1 \end{array} $\\
\hline $\;$ & $\begin{array}{c} 1g_{9/2} \\ 1g_{7/2} \end{array}$ & $ 6.6 \pm 0.6  $ & $ \begin{array}{c} 6.5 \\ 5.8 \end{array}$ & $ \begin{array}{c} 10.1 \\ 10.1 \end{array} $ & $-$ & $5.1 $ & $ 5.6$ & $ \begin{array}{c} 7.1 \\ 6.9 \end{array} $ & $ \begin{array}{c} 7.5 \\ 5.8 \end{array} $\\
\hline 
\end{tabular}
}\end{center}
\end{table}
\newpage


The present calculations 
(column FKVW) are evidently in very good agreement with data,
and comparable in 
quality or superior (especially for heavier hypernuclei) to the energy spectra 
calculated with other approaches. 
The QMC model
in its original form included a phenomenogical spin-orbit correction and the Pauli-blocking effect
at the quark level. Without these corrections the resulting energy levels show a 
strong overbinding (cf. Tab. 4 in Ref. \cite{Tsushima_old}). 
A very recent improvement \cite{Tsushima_new} solved the overbinding problem, introducing the scalar polarizability of the nucleon in a self-consistent way instead of the Pauli blocking correction. 
In Tables \ref{tab1} and \ref{tab2} we have included the latest update of these calculations.


\begin{figure}[h]
\begin{center}
\vspace{1.3cm}
\includegraphics[scale=0.55,angle=0]{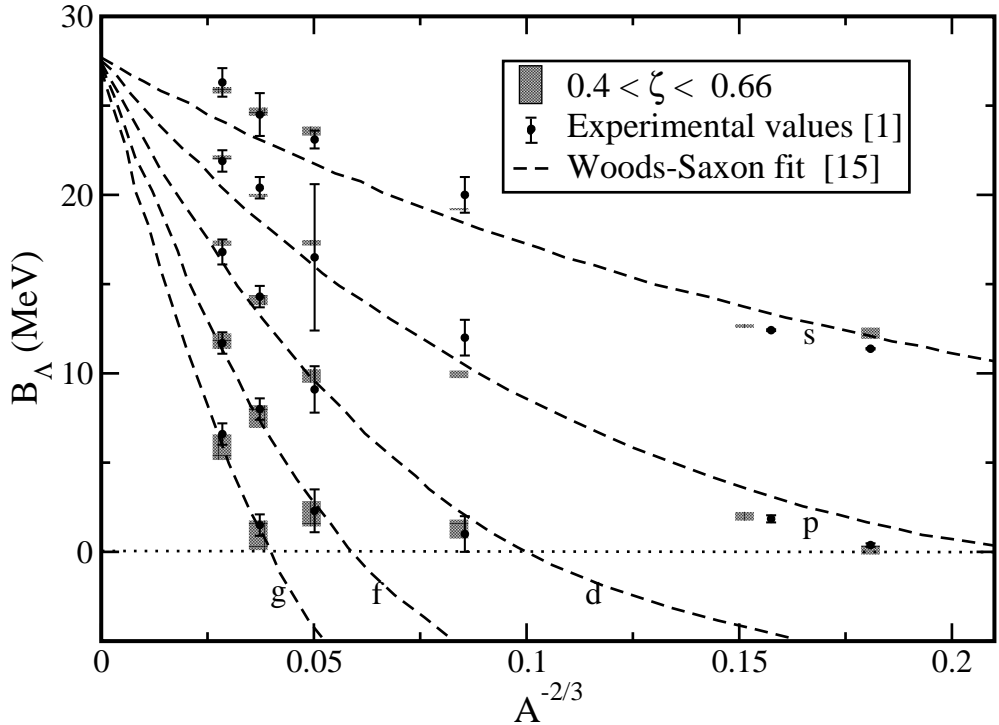}
\caption{\label{figB} Binding energies of the $\Lambda$  in different ($s,p,\dots$) orbitals of six hypernuclei (cf. Tables 
 \ref{tab1} and  \ref{tab2}),
calculated with the FKVW density functional using the three parameter sets 
for the $\Lambda N$ couplings (cf. Table \ref{tab0}). Results are plotted
as functions of the mass number and compared with experimental 
energies \cite{hypspinorbit}. Also shown is a Woods-Saxon fit \cite{MDG} (dashed curves) to guide the eye.
}
\end{center}
\end{figure}


Figure \ref{figB} provides a further test of the sensitivity of calculated single-$\Lambda$ 
energies with respect to a variation of the ratio $\zeta = G_{S,V}^{\Lambda (0)}/ G_{S,V}^{(0)}$
between contact terms representing the in-medium condensate background fields for the hyperon and the nucleons.  For the six hypernuclei 
listed in Tables \ref{tab1} and  \ref{tab2}, the $\Lambda$ binding energies 
calculated with the FKVW parameters plus the three best-fit parameter sets 
from Table \ref{tab0} that determine the $\Lambda N$ couplings, are plotted 
as functions of the mass number and compared with empirical energies. 
Calculations with all three parameter sets of the $\Lambda N$ interaction 
reproduce the data with high accuracy for a wide range of hypernuclear 
masses, using a reasonable band width of $\zeta$ values between 
the QCD sum-rule estimate ($\zeta\sim 0.4$) and the naive quark-model 
($\zeta =2/3$). 

In Tab. \ref{tab3}, we show the theoretical spin-orbit splittings for $\Lambda$ {\it p}-levels in comparison with
empirical values (second column) and other relativistic calculations with (fourth column) and without (fifth column)
tensor coupling.   

Concerning the $\Lambda$-nuclear spin-orbit splittings, these calculations have all been performed using the chiral SU(3) two-pion exchange $\Lambda N$ interaction with just the $\Sigma$ hyperon in the intermediate state. Adding terms with decuplet intermediate states as studied in Ref. \cite{camalich}
do not change the picture in any significant way, except that, looking at $\Delta \epsilon^\Lambda(p)$, there is then a preference for the smaller ratio $\zeta\simeq 0.4$, consistent with the in-medium QCD sum  rule analysis.\\

\begin{table}[h]
\caption{\label{tab3} P-shell spin-orbit splittings $\Delta \equiv\Delta \epsilon^\Lambda (p)$ for six hypernuclei ($^{13}_{~\Lambda}$C, $^{16}_{~\Lambda} $O,
$^{40}_{~\Lambda }$Ca, $^{89}_{~\Lambda }$Y, $^{139}_{\;\; \Lambda} $La, $^{208}_{\;\; \Lambda} $Pb). Experimental values \cite{929}, or empirical estimates \cite{hypspinorbit, Motoba016, y89exp}, are shown in comparison with our theoretical predictions (FKVW), using a broad range of $\zeta$ parameters (see Eq. \ref{zeta}), and other relativistic calculations with (RMFI \cite{tensor}) or without (RMFII \cite{Keil:1999hk}) tensor coupling. All energies are given in keV. The asterisk means that a local fit has been necessary.
}
\vspace{.3cm}
\begin{center}
{\small
\begin{tabular}{|cc||ccc|}
\hline
Nucleus & Exp. & FKVW & RMFI \cite{tensor} & RMFII \cite{Keil:1999hk}\\
&  $\Delta$~[keV] & $~(0.4 \le \zeta \le 0.66)$ & & \\
\hline\hline
$^{13}_{~\Lambda}$C & $152 \pm 54 \pm 36$ \cite{929}& $-160 \le \Delta \le 510$  & 310 & $\sim 1100^*$ \\\hline
$^{16}_{~\Lambda} $O & 
\begin{tabular}{c}
$300 \le \Delta \le  600$ \cite{Motoba016} \\
$-800 \le \Delta \le  200$ \cite{hypspinorbit}
\end{tabular}
%
%
& $-210 \le \Delta \le 490$ & 270& $\sim 1400$\\\hline
$^{40}_{~\Lambda }$Ca & $-$ &$ -140\le \Delta \le 420$ & 210 &$\sim1400$ \\ \hline
$^{89}_{~\Lambda }$Y & 90 ~\cite{y89exp} &$ -40\le \Delta \le 180$ & 110 & $\sim 700$\\ \hline
$^{139}_{\;\; \Lambda} $La & $-$ &$-20 \le \Delta \le 80$ & 50 & $\sim 300$\\ \hline
$^{208}_{\;\; \Lambda} $Pb & $-$ &$-20 \le \Delta \le 70$ & 50 & $\sim 300$\\\hline
\end{tabular}
}\end{center}
\end{table}


\section{Summary and conclusions}

1. A previously derived (FKVW) relativistic nuclear energy density functional, with constraints from
low-energy QCD, has been generalized to hypernuclei. In-medium chiral SU(3) dynamics is implemented at three-loop order in the energy density, with explicit treatment of two-pion (and kaon)
exchange $\Lambda$N interactions in the presence of the filled Fermi sea of nucleons.

2. Strong scalar and vector fields experienced by the $\Lambda$ are considered to emerge from its coupling to density-dependent quark condensates constrained by in-medium QCD sum rules. They are manifest in the form of contact terms that can be related to low-energy constants in a (chiral) effective field theory. The corresponding scalar and vector Hartree potentials have opposite signs and cancel in the central mean field. Their magnitudes are about 0.4 - 0.5 of the corresponding
mean fields for nucleons in nuclei. These reduced mean fields are consistent with the (admittedly uncertain) QCD sum rule analysis for nucleons and hyperons in a nuclear medium.

3. A $\Lambda$-nuclear surface coupling term, that appears in the
gradient expansion of a density functional for finite systems,
is generated model-independently from in-medium chiral SU(3) perturbation theory at the two-pion exchange level. This term proves to be important in
obtaining good overall agreement with $\Lambda$ single-particle spectra throughout the hypernuclear mass table.

4. The chiral two-pion exchange $\Lambda$N interaction in the presence of the nuclear core generates a (genuinely non-relativistic, model-independent) contribution to the $\Lambda$-nuclear spin-orbit force. This longer range contribution counterbalances the short-distance spin-orbit terms that emerge from scalar and vector mean fields, in just such a way that the resulting spin-orbit splitting of $\Lambda$ single particle orbits is extremely small. A three-body spin-orbit term of Fujita-Miyazawa type that figures prominently in the overall large spin-orbit splitting observed in ordinary nuclei, is absent for a $\Lambda$ in hypernuclei, simply because in a single-$\Lambda$ hypernucleus there is no Fermi sea of hyperons.

5. The confrontation of this highly constrained approach with empirical $\Lambda$ single-particle spectroscopy turns out to be quantitatively successful, at a level of accuracy comparable to that of the
best existing hypernuclear many-body calculations. The resulting $\Lambda$-nuclear single-particle
potential has dominant Hartree term with a central depth of  about $-30$ MeV, consistent with earlier
phenomenology.

6. While a relativistic framework has been used here for practical convenience, this is not mandatory. Given the cancellation of relativistic scalar and vector terms in the central mean field, their coherent effect in building up part of the spin-orbit force can be translated into derivatives of contact terms at next-to-leading order in an equivalent, non-relativistic effective field theory. The important compensating two-pion exchange mechanism that renders the overall spin-orbit coupling for the $\Lambda$ hyperon so abnormally small, is entirely of non-relativistic origin, as well as the similarly important surface gradient term that adds to the $\Lambda$-nuclear potential. \\

{\it Acknowledgements}. We thank Avraham Gal for numerous helpful 
discussions and a careful reading of the manuscript, and Jiri Mares for providing unpublished results. 
This research is partly supported by 
BMBF, GSI, INFN, MZOS (project 1191005-1010) 
and by the DFG cluster of excellence Origin and Structure of the Universe.
D. V. and P. F. are also supported by the {\it Agreement for Scientific
and Technological Cooperation between Italy and Croatia}.
P. F. acknowledges the kind hospitality of Marcella Grasso and the IPN staff at Orsay where part of this work has been done. 


{\appendix

\section{\label{Sec1App}Appendix: Gradient term of the $\Lambda$-nuclear 
density functional}

We briefly outline the chiral EFT estimate for the strength parameter 
$D_S^\Lambda$ of the gradient term in the interaction part of the 
$\Lambda$-nuclear density functional Eq.~(\ref{E_lambda_int}).

This parameter can be evaluated from the 
spin-independent part of the self-energy of a $\Lambda$-hyperon interacting
with weakly inhomogeneous nuclear matter.
Let the $\Lambda$ scatter from initial momentum $-{\bm q}/2$ to 
final momentum $+{\bm q}/2$, as shown in the two-pion exchange 
diagram of Fig. \ref{FigFeyn}. For small
momentum transfer ${\bm q}$ to the inhomogeneous nuclear medium, the
$\Lambda$ mean-field potential is modified by a correction term proportional to
${\bm q}^2$. In the energy density functional the factor ${\bm q}^2$ generates
(via Fourier transform) the product of density gradients: ${\bm \nabla}
\rho_\Lambda \cdot {\bm \nabla} \rho$ (with $\rho =2k_F^3/3\pi^2$ the nuclear
density).


\begin{figure}
\begin{center}
\includegraphics[scale=0.7,clip]{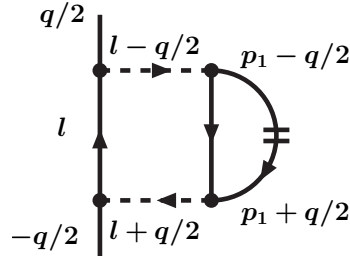}
\end{center}
\caption{\label{FigFeyn} Two-pion exchange between a $\Lambda$-hyperon and nucleons, with a 
$\Sigma$-hyperon and Pauli blocking effects in the intermediate state.
The horizontal double lines represent the filled Fermi sea of nucleons.}
\vspace{1cm}
\end{figure}


Consider now the two-pion exchange between the $\Lambda$ and the
nucleons in the medium. The small momentum transfer ${\bm q}$ enters both the
pion-baryon vertices, and the pion propagators. Expanding the diagrammatic
expression for the $\Lambda$ self-energy up to order ${\bm q}^2$, one
finds:
\begin{eqnarray} \rho D^{(\pi)}_\Lambda(\rho) & = & {D^2 g_A^2 M_B \over 6  f_\pi^4}
\int
\limits_{|{\bm p}_1| < k_F} {d^3 p_1 \,d^3 l\over (2\pi)^6}{{\bm l}^2\,(4\,{
\bm l}^4
+9\, {\bm l}^2\, m_\pi^2 +3\, m_\pi^4)\over (m_\pi^2 +{\bm l}^2 )^4\,
[\Delta^2 +
{\bm l}^2 -{\bm l}\cdot {\bm p}_1]} \nonumber \\
& \, &  \hspace{5.5cm}+ {\rm \;Pauli \; blocking \; terms}\,,
\label{eqApp1}
\end{eqnarray}
with the axial vector coupling constants $D=0.84$ and $g_A=1.3$. Here $M_B=1047\,
$MeV denotes an average baryon mass, and the $\Sigma\Lambda$ mass-splitting
$M_\Sigma - M_\Lambda = \Delta^2/M_B=77.5\,$MeV has been rewritten in
terms of the small scale parameter $\Delta = 285\,$MeV. 
The function $D^{(\pi)}_\Lambda(\rho)$ determines the (possibly density-dependent) strength of the gradient term. The
Pauli blocking correction in Eq. (\ref{eqApp1}) is obtained by reversing
the sign, and substituting the loop momentum as ${\bm l} = {\bm  p}_1 -  {\bm p}_2$ 
with $ |{\bm  p}_2 |< k_F$.  In this way the Pauli-blocked intermediate nucleon
states in the filled Fermi sea get properly removed from the loop integral.
Note also that the loop integral in Eq. (\ref{eqApp1}) is finite as it stands and 
does not require any regularization. It could even be solved in terms of elementary
functions (an arctangent and square-roots).

The density-matrix expansion of Negele and Vautherin \cite{NegeleVau} 
contributes a term proportional to ${\bm \nabla}^2 \rho$ to the in-medium insertion 
for an inhomogeneous many-nucleon system. Taking this term into account one 
obtains the following  additional contribution to the strength function:   
\begin{eqnarray} 
\rho D^{(\pi)}_\Lambda(\rho) & = & -{D^2 g_A^2 M_B \over 48
f_\pi^4k_F^4}\int
\limits_{|{\bm  p}_1| < k_F} {d^3 p_1 \,d^3 l\over (2\pi)^6}{35\,{\bm
l}^4\,(5\,{\bm p}_1^2
-3\,k_F^2) \over (m_\pi^2 +{\bm l}^2 )^2\, [\Delta^2 +{\bm l}^2
- {\bm l} \cdot
{\bm p}_1]} \nonumber \\
& \; & \hspace{5.5cm} +{\rm \; Pauli \; blocking \; terms} 
\label{eqApp2}
\end{eqnarray}
This combined loop and Fermi sphere integral is also convergent as it
stands. The weighting factor $5{\bm p}_1^2 -3k^2_F$ ensures that the 
divergent constant from the loop integral (scaling e.g. with an ultraviolet cutoff)
disappears in the final result.

\begin{figure}
\begin{center}
\includegraphics[scale=0.35,clip]{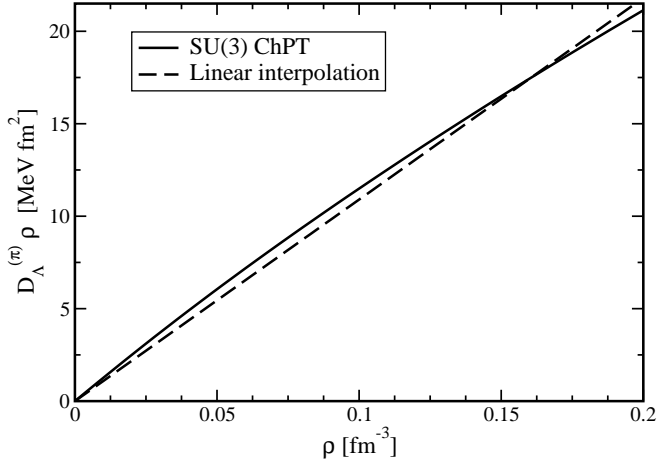}
\end{center}
\caption{\label{FigLambdader} The strength function $D^{(\pi)}_\Lambda(\rho) \rho$ related to the gradient term of the $\Lambda$-nuclear interaction, as function of the nucleon density $\rho$.}
\vspace{1cm}
\end{figure}

Summing up all four contributions from Eqs. (\ref{eqApp1}) and (\ref{eqApp2}), 
the resulting density-dependent strength function $\rho D^{(\pi)}_\Lambda(\rho)$ 
is shown in Fig. \ref{FigLambdader}. Since the density dependence is 
almost linear, the function $D^{(\pi)}_\Lambda(\rho)$ can be approximated 
by its value at nuclear matter saturation
density  $\rho_0 = 0.16\,$fm$^{-3}$~:
\begin{equation} 
\label{Dpi}
D^{(\pi)}_\Lambda(\rho_0)  = (139.3-21.0-10.6+1.3)\,{\rm
MeV}~{\rm fm}^5= 109\,
{\rm MeV~fm}^5\,. \end{equation}
Note that this estimate for the strength of the gradient term is a
model-independent and parameter-free result derived from the 
long-range $2\pi$-exchange with a $\Sigma$-hyperon in the intermediate state. 
Comparing the gradient term for the $\Lambda$-hyperon with
the corresponding term in $\Lambda$-nuclear density functional Eq. (\ref{der}), 
we obtain the following relation:
\begin{equation}
-D^{(\pi)}_\Lambda = D_S^\Lambda \; .
\end{equation}
From Eq. (\ref{Dpi}) it then follows:
\begin{equation}
D_S^\Lambda = -0.56 \; {\rm fm}^4 \;,
\end{equation}
remarkably close in value to the chiral EFT prediction for the strength 
of the corresponding gradient term in the nucleon sector of the density 
functional:  $D_S = -0.7$ fm$^4$ . 
}

\end{document}